\documentclass{article}

\usepackage[english]{babel}

\usepackage[letterpaper,top=2cm,bottom=2cm,left=3cm,right=3cm,marginparwidth=1.75cm]{geometry}

\usepackage{aas_macros}
\usepackage{amsmath}
\usepackage{graphicx}
\usepackage[colorlinks=true, allcolors=blue]{hyperref}
\usepackage{amssymb}	
\usepackage{subfigure}
\usepackage{natbib}
\usepackage{authblk}
\usepackage{xcolor}

\definecolor{ForestGreen}{RGB}{20,80,20}  

  {\endgroup}                      
  
\title{On the astrophysical origin of cosmic rays: Constraining the Ultra-High-Energy Cosmic Ray Horizon through Nearby Galaxy Distributions}
\author[1]{F., Dávila-Kurbán}
\author[2]{F. Duplancic}
\author[1,3]{D. Garcia Lambas}
\affil[3]{ \small{ \it
Observatorio Astronómico de Córdoba, UNC, Córdoba, Argentina.
}
}
\affil[1]{ \small{ \it
Instituto de Astronom\'\i{}a Te\'{o}rica y Experimental (IATE), CONICET, 
\hspace{6cm}
Observatorio Astron\'{o}mico de Córdoba,
Laprida 854, X5000BGR, C\'{o}rdoba, Argentina.
}
}
\affil[2]{ \small{ \it
Gabinete de Astronomía Extragaláctica, Departamento de Geofísica, Astronomía y Física, CONICET, Facultad de Ciencias Exactas, Físicas y Naturales, Universidad Nacional de San Juan, Av. Ignacio de la Roza 590 (O), J5402DCS, Rivadavia, San Juan, Argentina. 
}
}
\begin{document}
\maketitle

\begin{abstract}
The origin of ultra--high--energy cosmic rays (UHECR) remains one of the central open problems in astroparticle physics.
The observed large--scale anisotropy of UHECR arrival directions above several EeV, together with the suppression of the energy spectrum at the highest energies, suggests that their sources are extragalactic and that only a limited volume of the nearby Universe contributes significantly to the observed flux.
However, the effective spatial scale over which the local large--scale structure contributes to the observed anisotropy remains poorly constrained by current observations.
%
%
In this work we aim to constrain the effective propagation horizon of UHECR with energies \mbox{$E\geq8$~EeV} by studying how well the observed large-scale anisotropy is reproduced by the distribution of nearby galaxies at different distances, accounting for Galactic magnetic-field deflections.
%
%
We construct volume--limited galaxy samples from the GLADE+ catalogue extending to 10000~km~s$^{-1}$ ($\sim150$~Mpc) and use both dipole comparisons and angular cross--correlation analyses with the Pierre Auger Observatory UHECR flux map of 8~EeV and above to characterize the distance dependence of the anisotropy signal.
To assess the impact of Galactic magnetic deflections, we further weight the galaxy--UHECR correlations using deflection maps derived from a Galactic magnetic field model.
%
%
We find that the angular separation between the galaxy dipole and the UHECR dipole reported by the Pierre Auger Observatory is minimized for galaxies within $cz \lesssim 4000$~km~s$^{-1}$ ($\lesssim60$~Mpc), where the galaxy dipole amplitude is also maximal.
Independently, the galaxy--UHECR cross--correlation signal is dominated by this nearby population and rapidly weakens at larger distances.
Accounting for Galactic magnetic field deflections enhances the correlation amplitude by more than a factor of two, indicating that magnetic effects play a significant role in shaping the observed anisotropy.
%
%
Our results indicate that the observed large--scale anisotropy of UHECR of 8~EeV and above is primarily driven by sources within $\sim50$--60~Mpc.
This characteristic scale is consistent with expectations from energy losses and magnetic deflections, indicating that the observed anisotropy retains a measurable imprint of the nearby extragalactic matter distribution.

\end{abstract}

\section{Introduction}



The origin and nature of ultra-high-energy cosmic rays (UHECRs)—particles reaching the Earth with energies ranging from $\sim 1$ to over a hundred EeV ($1\text{ EeV} \equiv 10^{18}\text{ eV}$)—remain among the most intriguing puzzles in modern astrophysics. Observations from leading hybrid detectors have revealed that the mass composition becomes progressively heavier as energy increases \citep[][]{AugerSpectrum2020}.

The production and acceleration of such heavy, highly energetic nuclei within a Galactic scenario faces severe theoretical difficulties, as there are no reliable Galactic models to account for them. Consequently, there is a strong consensus that the most probable origin of UHECRs is linked to extragalactic sources located outside of the Galactic plane \citep{PierreAuger:2017pzq}. However, tracking these particles back to their sources is highly non-trivial: Galactic and extragalactic magnetic fields significantly deflect UHECR trajectories \citep{AlvesBatista2017, Farrar2014, Unger2024, KST2025}. This deflection depends directly on the magnetic rigidity of the particle, $R$, defined as the ratio between its energy and its nuclear charge ($Z$). Given that the origin and propagation of UHECR remain unclear, understanding the extragalactic distance limits of UHECR emitters is crucial to understand their astrophysical nature. 

The Greisen--Zatsepin--Kuzmin (GZK) effect \citep{Greisen1966, Zatsepin1966} establishes a fundamental energy-dependent limit to the propagation of UHECR. When particles exceed energies of $\sim 6 \times 10^{19}\,\mathrm{eV}$, they interact with photons of the cosmic microwave background (CMB), leading to significant energy losses. For protons, the dominant channel is photo-pion production, whereas heavier nuclei predominantly undergo photodisintegration processes. These interactions result in a steep suppression of the cosmic ray flux limiting the  UHECR propagation. This limit, commonly referred as the GZK horizon, typically ranges from several tens of Mpc for protons to only a few Mpc for intermediate and heavy nuclei at the highest energies. For example, while protons can reach the Earth from distances up to $\sim 100$ Mpc at energies around $6 \times 10^{19}\,\mathrm{eV}$, helium and carbon nuclei have horizons below $\sim 20$ Mpc, and iron nuclei are confined to even smaller scales due to their higher cross sections for photodisintegration.
Alternatively, this prominent spectral feature may not be exclusively driven by extragalactic propagation losses. It can be interpreted as an intrinsic property of the injectors, reflecting the maximum acceleration limits induced by the sources where the maximum energy scales with $Z$ under electromagnetic acceleration \citep{PierreAuger:2016_CombinedFit}. Such a scenario is compatible with hard injection spectra 
originating from relativistic magnetic reconnection \citep{Guo2014_MagneticReconnection} or unipolar induction in young pulsar magnetospheres \citep{Blasi2000_NeutronStarWinds, Kotera2015_PulsarEnvironment}. Furthermore, a population of sources with a variety of limits on their associated UHECR maximum rigidities was proposed to naturally synthesize a steep spectral cutoff and simulate an energy-dependent injection spectrum that shapes the observed flux suppression independently of the GZK horizon \citep{Kachelriess2006_FermiShock}.

Observational evidence has provided crucial validation of these theoretical predictions. The High Resolution Fly’s Eye (HiRes) experiment first reported the detection of the anticipated GZK suppression, observing a sharp flux reduction consistent with theoretical expectations at $E \approx 6 \times 10^{19}\,\mathrm{eV}$ \citep{Abbasi2008}. Subsequent analyses by the Pierre Auger Observatory and Telescope Array have confirmed this spectral feature while revealing additional complexity in the form of composition-dependent propagation effects \citep{Abraham2008,Abbasi2023}.

Beyond the basic picture, theoretical studies have emphasized the role of large-scale magnetic fields in shaping the effective UHECR horizon. The concept of ``magnetic horizons'' was investigated by \citet{Deligny2004}, showing that strong extragalactic magnetic fields can delay and isotropize cosmic ray trajectories, further reducing the effective contribution of distant sources even below the GZK threshold. 
Similarly, \citet{Harari2006} demonstrated that diffusion in turbulent magnetic fields can suppress the contribution of remote sources, yielding a sharper spectral cutoff than that expected from purely energy-dependent attenuation.
In their work, \citet{DeDomenico2013} investigated the influence of cosmology on the GZK horizon
of extragalactic ultra-high energetic protons, with energy ranging from 50 to 100 EeV. They considered different models of the Universe, from flat to curved ones, finding  non-negligible differences between the estimated values of the GZK horizon in Universes with different cosmological models.

Taken together, these results support a coherent framework in which the observed suppression at the highest energies is primarily shaped by the GZK effect, but modulated by additional astrophysical processes. In this context, the GZK horizon not only defines the maximum reach of UHECR-related astronomy, but also provides a unique probe of extragalactic environments and the composition of the most energetic particles in nature.

The search for correlations between ultra-high-energy cosmic rays (UHECR) and extragalactic sources has been a central topic in recent years. A significant dipolar anisotropy with \mbox{$E\geq8$~EeV} was found in Auger Collaboration data \citep{PierreAuger:2017pzq}. In that work, a Fourier analysis in right ascension reveals a dipole anisotropy with a 5.6$\sigma$ significance. Moreover, galaxies from the 2MASS Redshift Survey \citep[2MRS,][]{Huchra2012} has revealed a prominent dipole that differs by $55^\circ$ with the cosmic--ray flux direction. 


Despite these advances, the identification of UHECR progenitors is affected by important observational biases. In the case of the 2MRS the magnitude limit of the catalogue restricts the luminosity of galaxies at a given redshift (see, e.g., Fig~\ref{fig:cz_vs_MB}). Therefore, while bright sources can be observed at distances consistent with the GZK limit ($\sim$ 250 Mpc), faint galaxies are restricted to very close distances ($\sim$ 25 Mpc)  under--representing their actual contribution to the observed UHECR flux. 

It is therefore crucial to work with carefully defined and homogeneous galaxy samples that control for distance, luminosity, and galaxy type to isolate the underlying astrophysical processes responsible for UHECR acceleration. In this work, we employ such suitable galaxy control samples to study correlations with UHECR arrival directions. We specifically aim to constrain the maximum distance at which the dipole signal derived from the galaxy distribution deviates significantly from the observed UHECR dipole.
Furthermore, we study the roles of distance and the Milky Way magnetic field in the spatial correlation with UHECR, and will analyse the effect of more specific galactic properties in a follow-up study.
In Section 2, we describe the data catalogues employed throughout this work, namely the GLADE+ galaxy survey and the Pierre Auger Observatory UHECR data. 
We begin by calculating dipoles for both catalogues in Section 3. We perform cross-correlations between the catalogues to explore the effect of galaxy distance and the Milky Way magnetic field in Section 4. Finally, Section 5 presents a summary of the results and the conclusions.

\section{Data}
\label{data}

\subsection{Pierre Auger Observatory flux maps for $E \geq 8$~EeV}

The observational data used in this work are based on the all-sky flux maps of UHECR provided by the Pierre Auger Observatory. This observatory is designed for the indirect detection of UHECR through the extensive air showers they produce upon entering the Earth's atmosphere. These showers are recorded by an array of water-Cherenkov detectors distributed over a large area, complemented by fluorescence telescopes that measure the ultraviolet radiation emitted by atmospheric nitrogen excited by the shower particles.

The large collecting area of the observatory (approximately $3000~\mathrm{km}^2$) enables the accumulation of sufficient statistics at the highest energies, where the cosmic-ray flux is extremely low. Instead of analyzing individual arrival directions, we make use of the UHECR flux maps released by the Pierre Auger Collaboration for energies $E\geq8~\mathrm{EeV}$ \citep{PierreAuger:2017pzq}. These maps are constructed from 32187 detected events collected between January 2004 and August 2016 and represent the angular distribution of UHECR intensity on the sky after accounting for the detector exposure.

The resulting flux maps cover declinations in the range $-90^\circ < \delta < 45^\circ$, corresponding to approximately 85\% of the celestial sphere. This representation provides a statistically robust description of the large-scale anisotropy patterns in UHECR arrival directions and is particularly well suited for correlation studies with Galactic magnetic field deflection maps, where small-scale information from individual events is not required. Hereafter, we call this data \textit{Auger map}.

\subsection{Catalogues of extragalactic sources}

In this work, we used the extended version of the \textit{Galaxy List for the Advanced Detector Era} (GLADE+) galaxy catalogue \citep{Dalya2022}, an all–sky, homogenized compilation optimized for studies of the nearby large--scale structure. GLADE+ comprises about 23 million sources and was compiled from major photometric and spectroscopic surveys, such as Gravitational Wave Galaxy Catalogue \citep[GWGC,][]{White2011}, the Two Micron All Sky Survey \citep[2MASS,][]{Skrutskie2006},  HyperLEDA \citep{Makarov2014}, and the Sloan Digital Sky Survey quasar catalogue from the 16th data release \citep[DR 16Q,][]{Lyke2020}. The catalogue provides spectroscopic or photometric redshift information and distance estimates out to \mbox{$\sim$300} Mpc and is nearly complete in $B$ luminosity within $\sim$50 Mpc and remains highly complete for the most luminous galaxies out to $\sim$100$-$150 Mpc.
The plane of the Milky Way is masked to prevent incompleteness of galaxy catalogues in the Zone of Avoidance (ZOA). In this region the gas and dust of the Milky Way severely reduce the visibility of background galaxies. As documented in the GLADE+ description \citep{Dalya2022}, the exclusion of galaxies in this zone is necessary to avoid spurious inhomogeneities in the galaxy sky distribution in the regions close to the galactic plane.
%
The uniform sky coverage, with the exception of the Galactic plane, and depth of the catalogue allow us to probe the connection between nearby structures and the observed UHECR anisotropy at energies \mbox{$E\geq8$~EeV},  where the flux is expected to be dominated by sources within the GZK horizon.

\begin{figure}
    \centering
    \includegraphics[width=0.7\linewidth]{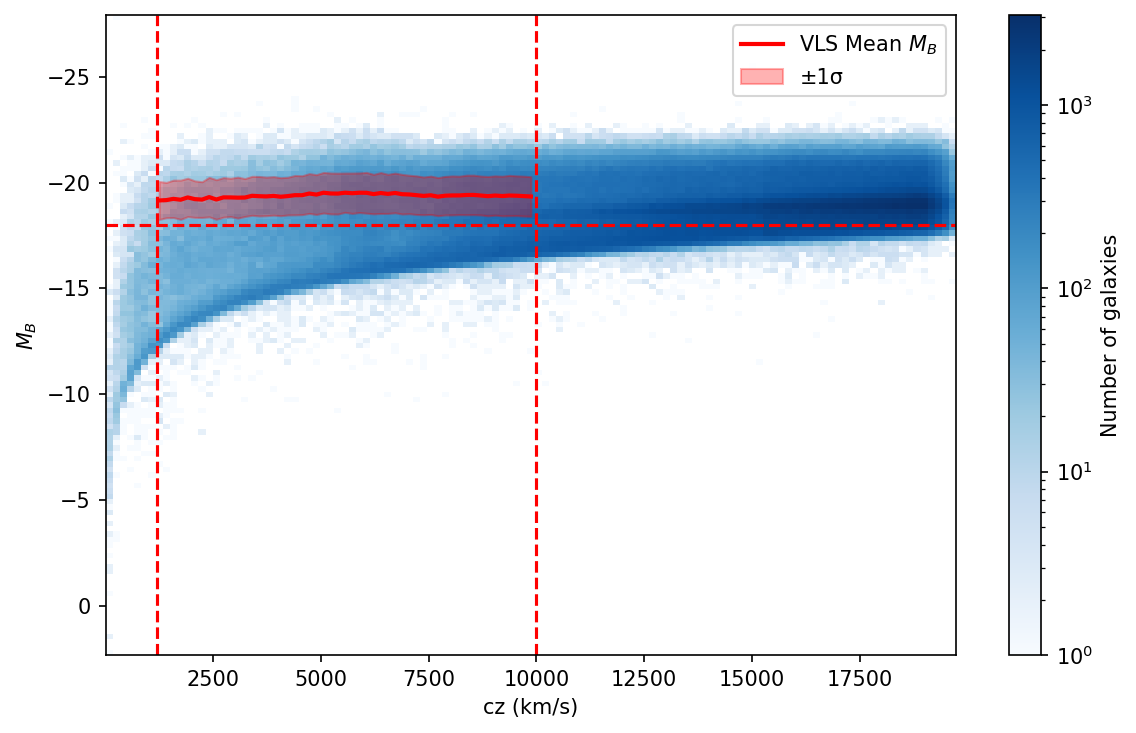}
    \caption{2D histogram of the radial velocity, $cz$, and B-band absolute magnitude, $M_B$, of the GLADE+ catalog. In red, the mean $M_B$ and $1\sigma$ deviation for the volume limited sample (VLS) chosen to ensure a constant density of galaxies across the distance range considered. The dashed lines indicate the cuts in $cz$ and maximum $M_B$ that define the GLADE+ VLS.}
    \label{fig:cz_vs_MB}
\end{figure}

To avoid biases arising from luminosity and distance differences, it is crucial to construct suitable galaxy samples. A volume-limited sample (VLS) includes all galaxies above a fixed luminosity threshold within a defined distance range, thereby ensuring uniform completeness across the surveyed volume and reducing distance-dependent selection biases. In this work we considered an absolute magnitude cut in the $B$ band of $M_B < -18$ which ensures an uniform trend of luminosity with distances in the range $1200~ \mathrm{km\,s^{-1}} < cz < 10000~\mathrm{km\,s^{-1}}$ (see Fig.~\ref{fig:cz_vs_MB}). These radial velocities correspond to luminosity distances of $18 < D_L <150~\mathrm{Mpc}\,h^{-1}$.
The lower distance limit was adopted to minimize the impact of galaxies whose redshifts are dominated by peculiar velocities, while the upper limit is consistent with theoretical expectations associated with the GZK horizon. Additionally, we consider a $\pm5\deg$ in galactic latitude to account for the Milky Way masking, and we exclude galaxies with declinations above $45^\circ$ for a better comparison with the declination restricted Auger data.

Under these assumptions, our final catalogue (hereafter GLADE+VLS) comprises $N=97761$ galaxies and provides a robust galaxy density field for quantifying large--scale anisotropies, constructing distance--binned dipole maps, and evaluating the expected matter-UHECR correlations under different magnetic deflection scenarios.

\begin{figure}
    \centering
    \includegraphics[width=1\linewidth]{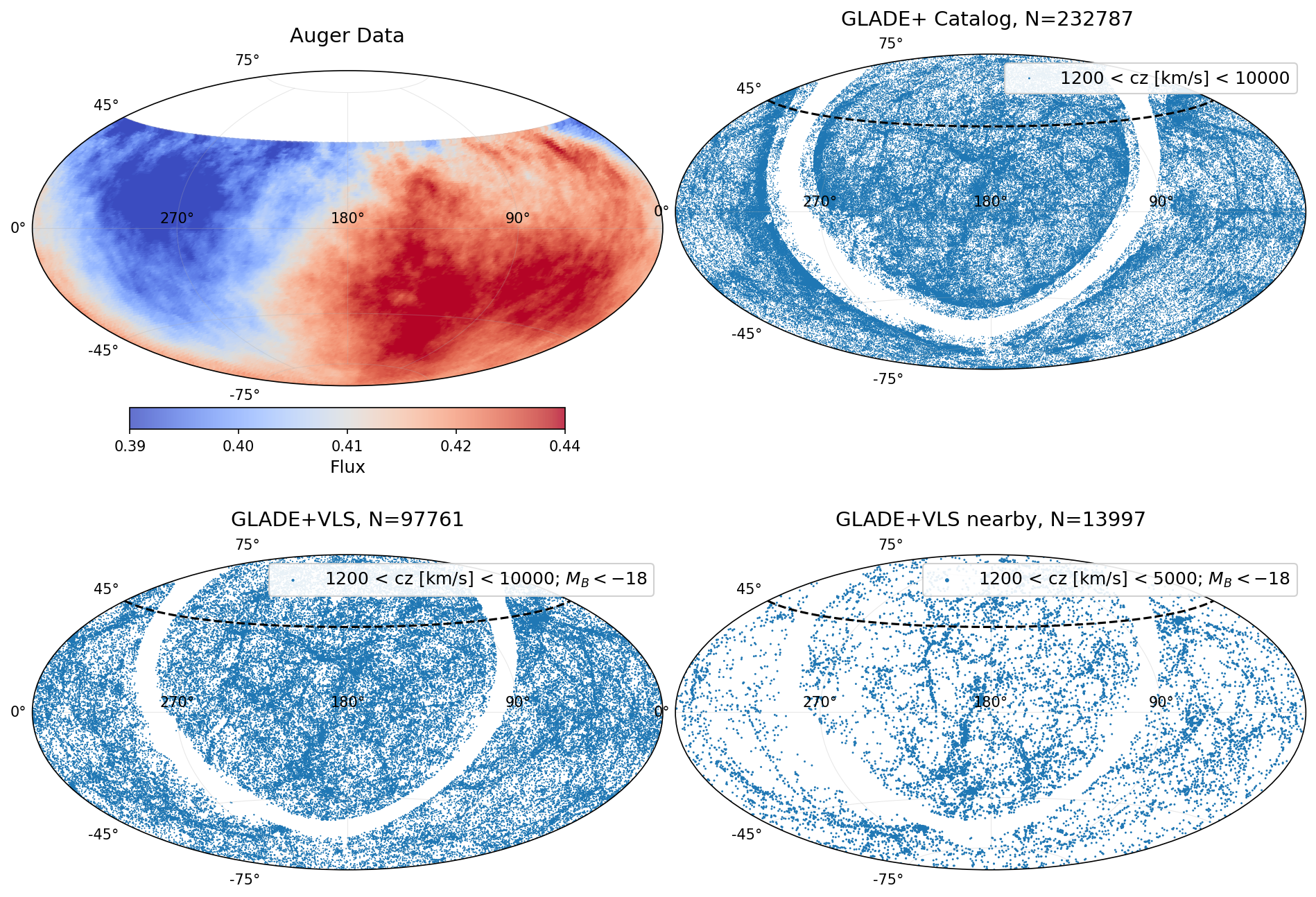}
    \caption{Top, sky-plots of the Auger flux map (left) and the GLADE+ galaxy catalogue within $1200~\mathrm{km\,s^{-1}}< cz < 10000~\mathrm{km\,s^{-1}}$ (right). Bottom, the bottom panels correspond to the GLADE+VLS (left) and a nearby subsample ($1200~\mathrm{km\,s^{-1}} < cz < 5000~\mathrm{km\,s^{-1}}$) of the GLADE+VLS.
    The blank stripe on the maps corresponds to the Milky Way plane masking of GLADE+ galaxy catalogue.
     The most nearby sample (bottom right) shows a significant asymmetry in the large-scale distribution of galaxies, consistent with the dipole in Auger data. 
     The black dashed line corresponds to $\delta=45^\circ$. Galaxies above this line are not considered in this study for a fair comparison with Auger data.}
    \label{fig:catalogos}
\end{figure}

The Auger flux map and the corresponding galaxy distributions from the GLADE+ catalogue are shown in Fig.~\ref{fig:catalogos}. The upper-left panel displays the UHECR flux with \mbox{$E\geq8$~EeV}. The dipolar modulation is clearly visible, with a broad excess and deficit defining the large-scale anisotropy.

The upper-right panel shows the full GLADE+ galaxy distribution within the velocity range $1200~\mathrm{km\ s^{-1}} < cz < 10000~\mathrm{km\ s^{-1}}$, comprising $N=232787$ galaxies, a distance range relevant for UHECR sources inside the GZK horizon. The resulting map traces the overall large-scale structure of the local Universe. In the lower-left panel, we plot the GLADE+VLS (restricted  galaxies with $M_B < -18$), comprising $N=97761$ sources. The cut enhances the contrast of the large-scale structure by suppressing low-luminosity systems that contribute mainly to shot noise, thereby providing a cleaner tracer of the underlying matter density field. Finally, the lower-right panel shows a more local subsample with  $1200~\mathrm{km\ s^{-1}} < cz < 5000~\mathrm{km\ s^{-1}}$ and $M_B < -18$ ($N=13997$). This nearby selection emphasizes the anisotropy of the matter distribution within $\sim70$ Mpc, where individual large-scale structures, as filaments, nodes and voids become more prominent. In this panel, a noticeable asymmetry is apparent, characterized by a relative deficit of galaxies at large right ascension values. This reflects the non-uniform distribution of nearby large-scale structures rather than an observational artifact.

The progressive restriction in distance and luminosity illustrates how the observed galaxy distribution transitions from a relatively smooth all-sky density field to a more anisotropic and structure-dominated pattern in the local volume. In the following sections, we explore whether this asymmetry in the nearby galaxy distribution is connected to the large-scale dipolar modulation observed in the Auger UHECR flux map.


\section{UHECR and galaxy dipoles}
\begin{figure}
    \centering
    \includegraphics[width=0.8\linewidth]{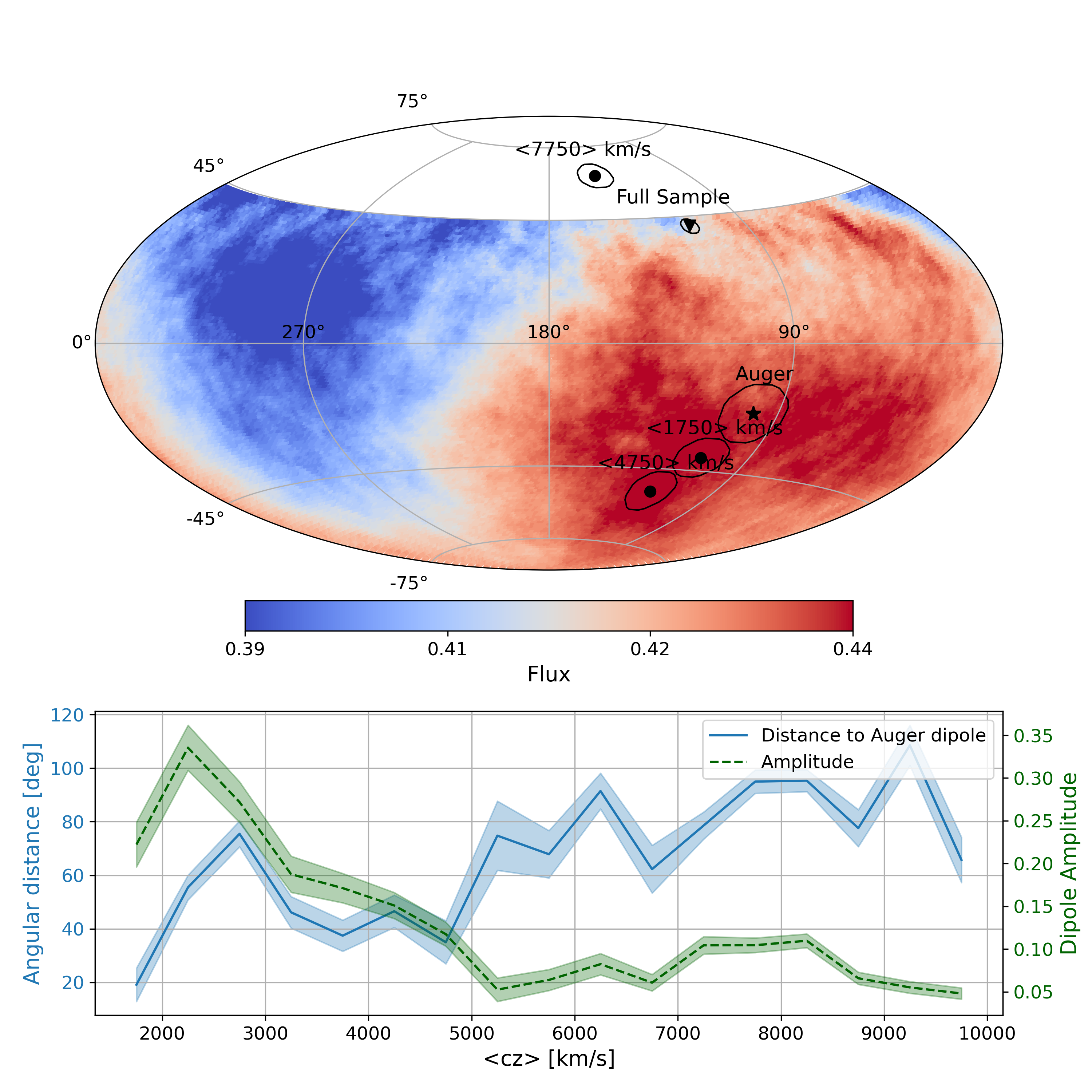}
    \caption{The top panel shows the Auger flux map with the dipoles of three representative GLADE+VLS subsamples plotted with black full circles and a circumference showing a $1\sigma$ bootstrap error estimation. The GLADE+VLS (full sample) dipole is represented with a black triangle. In the bottom panel, the galaxies dipole amplitude and distance to Auger dipole, vs. $cz$, where the shaded regions correspond to the $1\sigma$ bootstrap error estimation. Only the dipoles of nearby galaxies show high amplitude and are close to the Auger dipole.}
    \label{fig:dipoles_vs_cz}
\end{figure}

The calculation of a dipole is a broad tool used for detecting large-scale anisotropies of a given distribution. Here, we follow the study of \cite{PierreAuger:2017pzq} and perform a Rayleigh analysis, in right ascension, for our galaxy VLS sample.

The first harmonic Fourier components are calculated as:

\begin{equation}
    a_\alpha = \frac{2}{W} \sum_{i=1}^N w_i \cos\alpha_i,\;\; \
    b_\alpha = \frac{2}{W} \sum_{i=1}^N w_i \sin\alpha_i,
\end{equation}

\noindent where the sums take into account all galaxies with right ascension $\alpha_i$, and the weights $w_i$ are normalized in the factor $W=\sum_{i=1}^Nw_i$. These weights account for the ``flux'' or luminosity of each galaxy defined as $w_i=10^{-0.4*M_{B_i}}$.

Finally, the amplitude, $r_\alpha$, and phase, $\phi$, of the dipoles are given by:

\begin{equation}
    r_\alpha = \sqrt{a_\alpha^2+b_\alpha^2},\;\; \
    \tan\phi_\alpha = \frac{b_\alpha}{a_\alpha}.
\end{equation}


As a reference, we adopt the UHECR dipole coordinates and uncertainties reported by the Pierre Auger Collaboration \citep{PierreAuger:2017pzq}, specifically $(\alpha, \delta) = (100^\circ\pm10, -24^\circ\pm13)$. We confirmed these coordinates by applying our dipole estimator directly to the Auger flux map, recovering a consistent direction within reported uncertainties\footnote{An offset with respect to the published values is expected, as the flux map is a pixelized HEALPix representation rather than a list of individual events, and the mean flux is subtracted prior to computing the dipole to remove the monopole bias introduced by the partial sky coverage.}: $(\alpha, \delta) = (98^\circ, -21^\circ)$. The bootstrap uncertainties quoted in this section (500 iterations) correspond to the galaxy dipole directions, obtained by resampling the GLADE+VLS with replacement. This approach accounts for spatial variance in the galaxy distribution and yields more conservative uncertainty estimates than standard Poisson errors.

We perform the dipole calculation in successive slices of radial velocity $cz$ of the GLADE+VLS to investigate the evolution of the galaxy dipole. The results are presented in Fig.~\ref{fig:dipoles_vs_cz}. In the top panel, we display the dipole position for the full galaxy sample together with three representative subsamples with mean velocities of 1750, 4750, and $7750~\mathrm{km\,s^{-1}}$. The bottom panel shows the evolution of both the angular separation between the galaxy dipole and the Auger Collaboration dipole, as well as the corresponding galaxy dipole amplitude as a function of radial velocity.

We find that, as expected due to the large scale homogeneity of the Universe, the galaxy dipole decreases with distance. It is worth noticing that the prominent amplitude ($r_\alpha \sim 0.25 - 0.35$) obtained in these nearby bins is consistent in order of magnitude with independent calculations \citep[e.g.,][]{Erdogdu2006dipoleanisotropy2massredshift}. We notice moreover that the relative separation between the galaxy and the Auger dipoles increases with distance. At $cz < 5000~\mathrm{km\,s^{-1}}$ the mutual dipole angular separation is around 40$^\circ$ while at $cz> 5000~\mathrm{km\,s^{-1}}$ this separation is 80$^\circ$ on average. The smaller dipole angular separations for the nearby galaxy subsamples suggest a physical correlation with UHECR that contributes to the observed UHECR anisotropy.


It is worth noting that recent harmonic analyses of UHECR data have shown that, beyond the dipole component, higher multipole moments do not reach a pre-trial significance above 3$\sigma$ \citep{Tinyakov2022}, with the only robust detection corresponding to the dipole reported in \citet{PierreAuger:2017pzq}. This indicates that, with current statistics, the information content accessible through individual spherical harmonic coefficients is limited. In this context, complementary statistical approaches become particularly valuable.


%

%


\section{Cosmic Rays--Galaxy cross-correlation}

The results presented in the previous section show that the alignment between the galaxy dipole and the UHECR dipole weakens at radial velocities beyond $5000~\mathrm{km\,s^{-1}}$, while nearby galaxy samples exhibit both a larger dipole amplitude and a smaller angular separation with respect to the Auger Collaboration dipole. This behaviour suggests that the observed UHECR anisotropy is primarily influenced by the local large-scale structure.

While the dipole analysis provides a global characterization of large-scale anisotropies, it does not capture potential correlations on smaller angular scales. To further investigate the connection between UHECR and their candidate extragalactic sources, it is therefore necessary to explore the presence of angular cross-correlations between UHECR arrival directions and nearby galaxies.

To this end, we employ the angular cross-correlation function, a fundamental statistical tool for characterizing clustering in astronomy and cosmology. Originally introduced in studies of galaxy clustering \citep[e.g.,][]{Peebles1973ApJ,peebles1980large}, it quantifies the excess probability, relative to a random distribution, of finding two objects separated by a given angular distance. Because it operates directly in configuration space, the correlation function provides an intuitive description of anisotropies across angular scales and is particularly robust in the presence of incomplete sky coverage or limited statistics.

From a mathematical perspective, the angular correlation function $w(\theta)$ contains the same statistical information as the spherical harmonic power spectrum $C_\ell$. The two quantities are related through the relation \citep{peebles1980large}

\begin{equation}
w(\theta)=\sum_{\ell=0}^{\infty}\frac{2\ell+1}{4\pi}C_\ell P_\ell(\cos\theta),
\end{equation}

\noindent where $P_\ell$ are the Legendre polynomials. Thus, while the $C_\ell$ describe the anisotropy field in harmonic space, the correlation function provides an equivalent representation in configuration space, organizing the same information in terms of angular separations. This property makes this statistical tool particularly suitable for investigating physically motivated correlations between UHECR arrival directions and the nearby matter distribution.

If the sources of UHECR trace the large-scale distribution of matter in the local Universe, a positive cross-correlation signal is expected at angular scales comparable to the typical magnetic deflections experienced during propagation. In this sense, since UHECR above a few EeV are expected to originate predominantly within the GZK horizon, the anisotropy of their arrival directions should reflect the inhomogeneous distribution of nearby matter. Cross-correlating UHECR flux maps with galaxy catalogues therefore provides a physically motivated test of whether candidate sources follow the large-scale structure of the local Universe.

\subsection{Cross-correlation procedure}

We measure the angular count-scalar cross-correlation function between a set of discrete sources, $N$, and a scalar field, $F$, using the \texttt{NKCorrelation} estimator implemented in \textsc{TreeCorr} \citep{Jarvis2004}. In our case, $N$ represents the galaxy positions and $F$ corresponds to the Pierre Auger Observatory flux map.
For each angular separation bin $\theta$, the correlation is defined as the difference between the mean flux measured around the $N$ objects and the mean flux measured around a uniformly distributed random catalogue, normalized by the global mean flux of the $F$ map:

\begin{equation}
w_{NF}(\theta) =
\frac{\langle F \rangle_{\mathrm{data}}(\theta) - \langle F \rangle_{\mathrm{random}}(\theta)}{\langle F \rangle_{\mathrm{data}}},
\label{eq:wnk}
\end{equation}

\noindent  where $\langle F \rangle_{\mathrm{data}}(\theta)$ is the mean flux of the $F$ map at angular separation $\theta$ from the $N$ sources, $\langle F \rangle_{\mathrm{random}}(\theta)$ is the corresponding quantity measured around the random catalogue, and $\langle F \rangle_{\mathrm{data}}$ is the mean flux over the full sky region considered. This normalization ensures that $w_{NF}(\theta)=0$ when the local mean flux equals the global mean, with positive (negative) values indicating an excess (deficit) of flux at the given angular scale.

We estimate statistical uncertainties using the internal jackknife resampling method implemented in \textsc{TreeCorr}, based on a patch decomposition of the survey area. The number of patches is chosen so that each patch has an angular diameter of approximately $20^\circ$, ensuring adequate statistical sampling. We verify that each patch contains a sufficient number of objects to provide a robust variance estimate. Although this choice may slightly overestimate uncertainties on small angular scales, it allows us to account for both shot noise and cosmic variance in a conservative manner.

Unlike the dipole analysis, this estimator probes correlations on specific angular scales and is therefore sensitive to localized associations between galaxies and UHECR arrival directions. This statistic provides a direct and integrated measure of the joint angular distribution of galaxies and UHECR events over a wide range of angular scales. Its implementation does not require special treatment of sky regions without data, since the random catalogs are defined over the same sky coverage as the galaxy sample.
By comparing the cross-correlation functions obtained for different galaxy subsamples with the UHECR flux maps, we can assess the degree of association between cosmic-ray arrival directions and the properties of their candidate extragalactic sources.

\subsection{Effect of galaxy distance}


The dipole analysis presented in the previous section indicates that the alignment between the galaxy and UHECR dipoles is strongest for nearby galaxies. To further quantify the role of distance in the angular cross-correlation between galaxies and cosmic rays, we investigate how the correlation signal depends on the galaxy distance by dividing the GLADE+VLS into three velocity-distance bins: $1200$-$4000$, $4000$-$7000$, and $7000$-$10000~\mathrm{km\,s^{-1}}$.

\begin{figure}
    \centering
\includegraphics[width=1\linewidth]{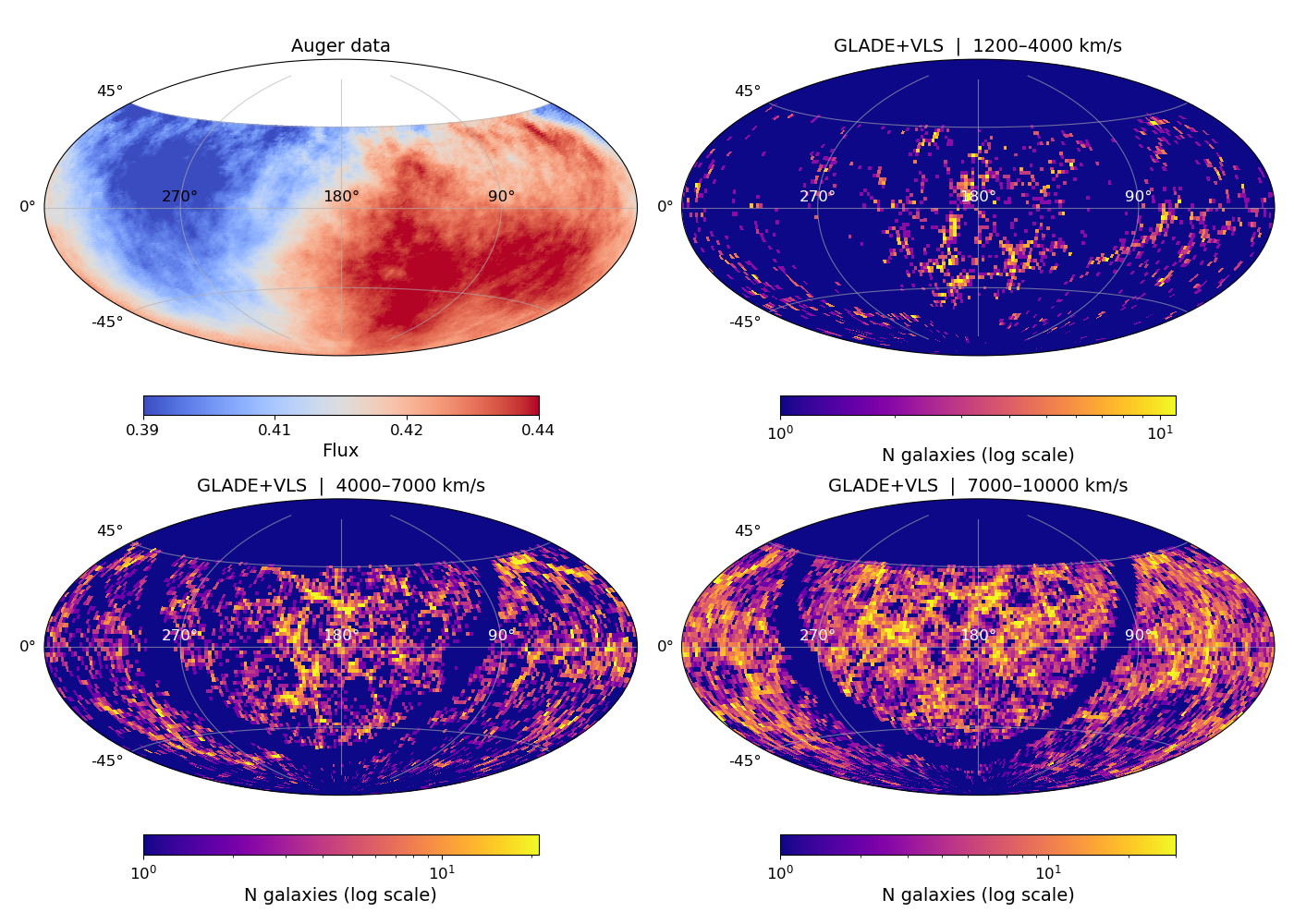}
    \caption{Sky-map of Auger flux (upper left), and different $cz$ subsamples of the GLADE+VLS used in the correlation analysis. A density map is shown for each of the three velocity-distance subsamples to highlight the prominent structures in each case.}
    \label{fig:skymap_cz}
\end{figure}

Fig.~\ref{fig:skymap_cz} shows the sky distribution of galaxies in each distance bin as well as the UHECR flux map. Nearby galaxies exhibit a more pronounced clustering pattern, while the distributions become progressively more homogeneous at larger distances, as expected from large-scale structure dilution and survey completeness effects.

\begin{figure}
    \centering
\includegraphics[width=1\linewidth]{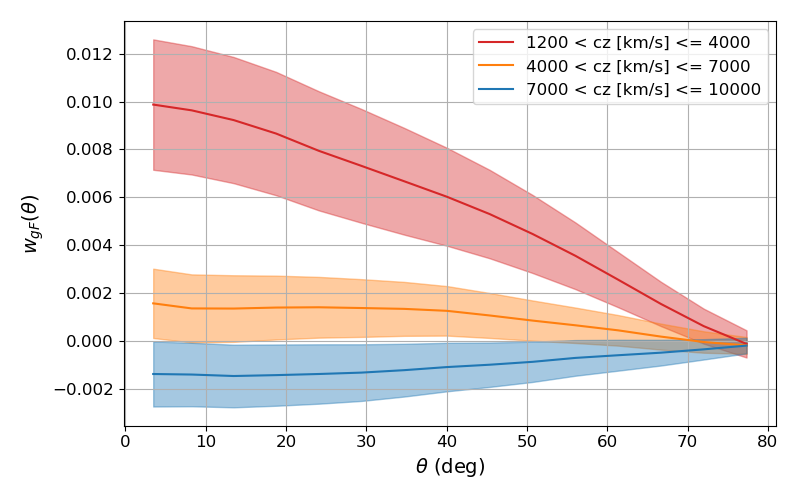}
    \caption{Cross-correlation functions between the Auger flux map and the three GLADE+VLS galaxy samples. Jackknife uncertainty estimations are represented as shaded bands.
    }
    \label{fig:corr_dist}
\end{figure}

The corresponding angular cross-correlation functions between the galaxy samples and the Auger flux map are presented in Fig.~\ref{fig:corr_dist}. The correlation signal is clearly dominated by the nearest galaxy bin ($cz < 4000~\mathrm{km\,s^{-1}}$), while the amplitude decreases significantly for the intermediate and farthest distance bins. This result indicates that the association between galaxy positions and UHECR arrival directions is primarily driven by the local galaxy population.

The choice of $cz = 4000~\mathrm{km\,s^{-1}}$ as a transition scale is further motivated by the dipole analysis discussed in the previous section, where a similar distance range exhibits both the largest dipole amplitude and the smallest angular separation with respect to the Auger Collaboration dipole. The consistency between the dipole and cross-correlation results strengthens the conclusion that the observed UHECR anisotropy is dominated by structures in the nearby Universe.

This behavior is naturally interpreted in the context of the GZK horizon, which limits the contribution of distant extragalactic sources to the observed UHECR flux. Together, these results support a scenario in which the angular distribution of UHECR is primarily shaped by the local large-scale structure traced by nearby galaxies.



\subsection{Magnetic field deflection model}
\label{sec:mf_model}

The cross-correlation analysis presented above is sensitive to the underlying association between nearby galaxies and the observed arrival directions of ultra-high-energy cosmic rays. To quantify the impact of magnetic deflections on these trajectories, we adopt particle rigidity -- defined as the ratio between the energy of the particle and its nuclear charge ($R = E/Z$) -- as the fundamental parameter determining the magnitude of its angular deviation. For any given magnetic-field configuration, particles with identical rigidity experience similar deflections independently of their mass or nuclear species. Measurements of the energy spectrum and mass composition by the Pierre Auger Collaboration indicate a transition toward an increasingly heavier composition approaching the flux suppression region, yielding characteristic rigidities of $R \sim 2-3~\mathrm{EV}$ for heavy nuclei, $R \sim 3-6~\mathrm{EV}$ for intermediate-mass components, and extensions up to $R \gtrsim 8~\mathrm{EV}$ for the lightest nuclei \citep{AugerComposition2014, AugerSpectrum2020}. These values are consistent with an electromagnetic acceleration paradigm where the maximum energy scales with charge, implying a common cutoff in rigidity. Based on this distribution, we adopt a representative rigidity of $R = 5~\mathrm{EV}$ to characterize the light-to-intermediate mass component that carries the bulk of the observed UHECR flux.

In the local Universe, where typical propagation distances are relatively short, deflections induced by extragalactic magnetic fields (EGMF) are expected to be at most a few degrees for rigidities $R \gtrsim 5~\mathrm{EV}$ \citep{Dolag2005, AlvesBatista2017}. Consequently, we neglect EGMF deflections at first order. 
To empirically validate this assumption against localized regions of higher magnetic
intensity, we performed a robustness test by masking the regions associated to the most nearby galaxy clusters catalogued by \citet{ReiprichBohringer2002}.
Specifically, within the region $cz < 4000~\mathrm{km\,s^{-1}}$ where our directional correlation peaks, the catalogue contains 10 clusters (including Virgo, Fornax, Antlia, Hydra, Centaurus and others).
For these systems, we adopt masks based on $R_{500}$, the radius within which the mean enclosed density reaches 500 times the critical density of the Universe. 
This radius traces the extent of the virialized, X-ray-emitting intracluster medium, providing a conservative estimate of the region over which cluster-associated magnetic fields could plausibly affect UHECR propagation. 
The angular projections of these physical boundaries for the 10 clusters cover 0.22\% of the Auger-covered sky (0.18\% of the full sky).
After removing all galaxies within these regions, the recalculated GMF-deflected cross-correlation functions showed no significant modifications.
This shows that the correlation signal of Fig.~\ref{fig:corr_dist} is neither driven nor biased by localized cluster environments.

In contrast, deflections produced by the Galactic magnetic field (GMF) can reach several tens of degrees over most of the sky at these rigidities \citep{Dolag2005, AlvesBatista2017}. Therefore, it is crucial to account for the GMF in our correlation study, modeling the angular smearing of UHECR arrival directions solely through the Galactic magnetic structure. To assess the robustness of our results against variations in this baseline rigidity, we additionally explore $R = 2~\mathrm{EV}$ and $R = 8~\mathrm{EV}$ in Appendix~\ref{appendix:A}. These alternative values cover the dominant rigidity range inferred from observations and probe, respectively, stronger and weaker magnetic-deflection regimes, thereby encompassing the full expected range of deflections for the nuclei contributing to the observed spectrum.

To model UHECR propagation in the GMF, we employed the CRPropa 3 framework \citep{AlvesBatista2016} and performed backtracking simulations of charged particles through the \citet[][hereafter UF23]{Unger2024} GMF model. Antiparticles were propagated from the observer position at Earth, assumed to be located at the solar position at a distance of $8.5$~kpc from the Galactic center, to the boundary of the Milky Way, modeled as a sphere of radius $20$~kpc. This radius encloses the region where the GMF is observationally constrained and where the dominant contribution to UHECR deflections is expected to arise. This procedure provides a direct mapping between arrival directions at Earth and their corresponding directions at the Galactic boundary.


The celestial sphere was discretized using a HEALPix grid \citep{Gorski2005} with $\mathrm{NSIDE}=64$, corresponding to $N_{\mathrm{pix}}=49152$ directions and an angular resolution of approximately $0.9^{\circ}$. For each direction, the angular deflection was computed as the separation between the injected direction at the observer and the exit direction at the Galactic boundary. This procedure yields an all-sky map of GMF-induced deflections at fixed rigidity.

The resulting deflection map was smoothed using a Gaussian kernel with a width of $40^{\circ}$, implemented through HEALPix utilities. This angular scale was chosen to match the effective resolution of the UHECR flux maps with energies $E\geq8$~EeV used in this work, which are dominated by large-scale anisotropy patterns rather than localized features. Smoothing over comparable angular scales suppresses pixel-scale noise and enhances coherent structures in the arrival direction distribution \citep{PierreAuger:2017pzq}.

Applying the same smoothing scale to the GMF deflection map ensures a consistent comparison between magnetic deflections and UHECR flux anisotropies in the subsequent correlation analysis. This approach prevents spurious correlations driven by small-scale fluctuations and maximizes sensitivity to large-scale directional modulations induced by the GMF.

\begin{figure}
   \centering
\includegraphics[width=0.75\linewidth]{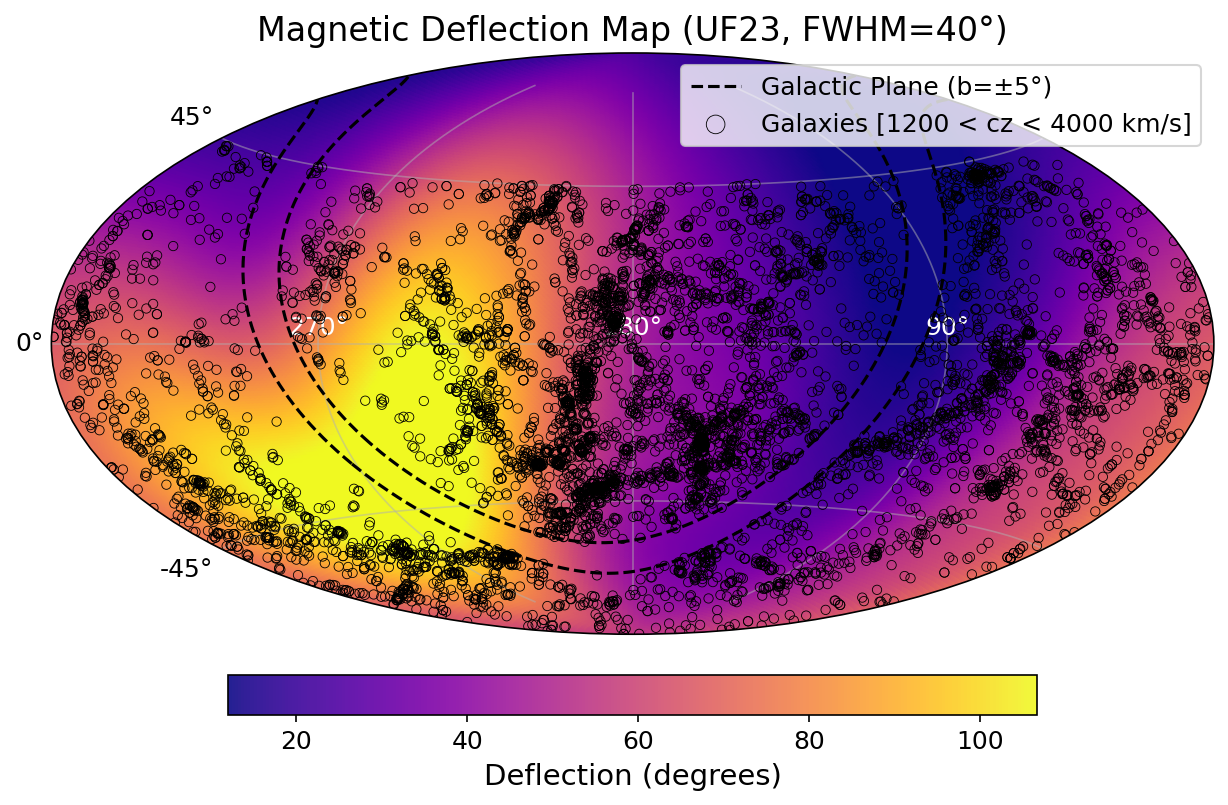}
   \caption{Skymap of Galactic magnetic deflection model of \citet[][UF23]{Unger2024} adopting a rigidity value of $R=5$~EV with a 40$^\circ$ smoothing, and GLADE+VLS nearby galaxies \mbox{($1200 < cz < 4000$)}. The galactic plane is delimited with black dashed lines.}
   \label{fig:skymap_defl}
\end{figure}

Fig.~\ref{fig:skymap_defl} shows the resulting smoothed GMF deflection map together with GLADE+ galaxies in the range $1200 < cz < 4000~\mathrm{km\,s^{-1}}$ and $M_B \leq -18$, marked as white circles. The region of maximum deflection is located toward the Galactic centre, approximately at $\alpha=266.4^{\circ}$ and $\delta=-29.0^{\circ}$.

\subsection{Effect of MW magnetic field deflections}

To investigate the impact of GMF on the cross-correlation between galaxies and UHECR arrival directions, we incorporate the GMF deflection map into the correlation estimator. Specifically, we modify the estimator in Eq.~\ref{eq:wnk} by assigning an additional weight to each galaxy--field pair that accounts for the local magnitude of magnetic deflections.

Each galaxy is weighted by $w_i = \theta_i^{-2} / \langle \theta^{-2} \rangle$, 
where $\theta_i$ (in degrees) is the angular deflection predicted by the GMF model 
at the position of the $i$-th galaxy and $\langle \theta^{-2} \rangle$ denotes the average 
over all galaxies in the sample.  The weights are normalized to have a mean of~1, 
which removes any dependence on the absolute unit of the deflection angle.  
This weighting scheme effectively accounts for the solid angle over which 
UHECR originating from a given direction are spread by Galactic 
magnetic deflections.


By weighting potential sources by the inverse of the deflection area, we enhance the contribution of galaxies located in regions of the sky where UHECR trajectories are less distorted by the GMF. If nearby galaxies trace the true sources of UHECR, incorporating GMF deflections in this manner is expected to increase the amplitude of the observed cross-correlation signal.

\begin{figure}
   \centering
\includegraphics[width=0.85\linewidth]{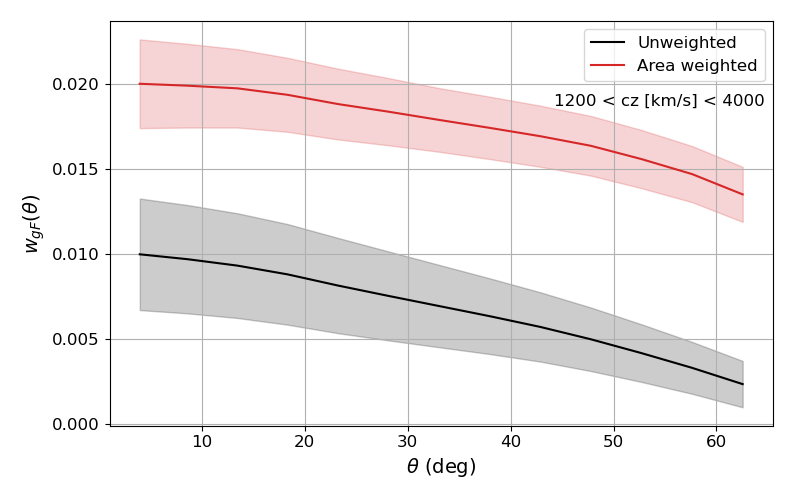}
   \caption{
   Weighted cross-correlation between the Auger flux map and GLADE+VLS nearby galaxies ($1200 < cz < 4000~\mathrm{km\,s^{-1}}$ and $M_B \leq -18$) using inverse GMF deflection-area weights for $R=5$~EV.
   Shaded bands represent Jackknife uncertainty estimations.}
   \label{fig:correlation_weights}
\end{figure}

The resulting weighted cross-correlation function is shown in Fig.~\ref{fig:correlation_weights}. When the inverse GMF deflection area is included, the correlation amplitude increases by more than a factor of two with respect to the unweighted case. We tested alternative weighting prescriptions (see Appendix~\ref{appendix:A}) and obtained qualitatively similar results; however, the deflection-area weighting maximizes the enhancement of the correlation signal.

This result supports the interpretation that GMF deflections play a significant role in shaping the observed UHECR arrival directions and that nearby galaxies remain the dominant contributors to the correlation signal once these effects are properly taken into account.

\section{Summary and conclusions}
\label{conclusions}

In this work we aim to constrain the effective propagation horizon of UHECR with energies \mbox{$E\geq8$~EeV} by studying both their large-scale anisotropy and its cross-correlation with the distribution of nearby galaxies. To achieve this goal, we employed homogeneous and carefully selected galaxy samples, constructed to minimize distance-dependent selection effects. This approach is essential to avoid spurious correlations and to ensure that the measured anisotropy signals genuinely reflect the underlying matter distribution in the local Universe.

Our dipole analysis shows that the angular separation between the galaxy and UHECR dipoles is minimized for the closest galaxy sample, which also exhibits the largest dipole amplitude. As progressively more distant galaxy populations are included, the dipole alignment weakens and the amplitude decreases. These results naturally motivate a focused investigation of the local Universe and indicate that the dominant contribution to the observed UHECR anisotropy arises from structures within $cz \leq 4000~\mathrm{km\,s^{-1}}$, including the local void.

To further quantify this association, we adopted an angular cross-correlation function approach. This methodology, widely used and extensively validated in large-scale structure and cosmological studies, provides a robust statistical framework to probe anisotropies on specific angular scales. Its application to UHECR data represents a novel aspect of this work, enabling a more detailed characterization of the relationship between cosmic-ray arrival directions and the distribution of potential extragalactic sources. The cross-correlation analysis confirms the dipole results, yielding a statistically significant signal only for the nearest galaxy sample, while no correlation is detected for more distant populations.

We incorporated the effect of Galactic magnetic field by introducing weighting schemes based on the predicted deflection angle and deflection area in the cross-correlation estimator. Accounting for these magnetic deflections leads to an enhancement of the correlation amplitude by more than a factor of two with respect to the unweighted case. By down-weighting sources located in regions of the sky subject to stronger magnetic deflections, our approach effectively optimizes the sensitivity to physically meaningful associations between galaxies and UHECR arrival directions.

Taken together, our results provide an observational determination of the effective UHECR source horizon, consistent with the theoretical GZK limit, and reinforce the interpretation of the observed anisotropy as originating predominantly in the local extragalactic Universe.

Ongoing and future work will extend this methodology to investigate the nature of the astrophysical progenitors of UHECR. By applying the same cross-correlation framework to specific galaxy populations and candidate source classes, we aim to further constrain the physical origin of the highest-energy particles in the Universe.

\appendix
\section{Additional Rigidity values and weighting schemes for GMF analysis}
\label{appendix:A}

To test the robustness of our results against the choice of rigidity values, we repeated the full Galactic magnetic field (GMF) analysis using values of $R = 2$ and $8$~EV. As mentioned in Sec.~\ref{sec:mf_model}, these values encompass the expected range for the composition of UHECR.

\begin{figure}
    \centering
    \includegraphics[width=1\linewidth]{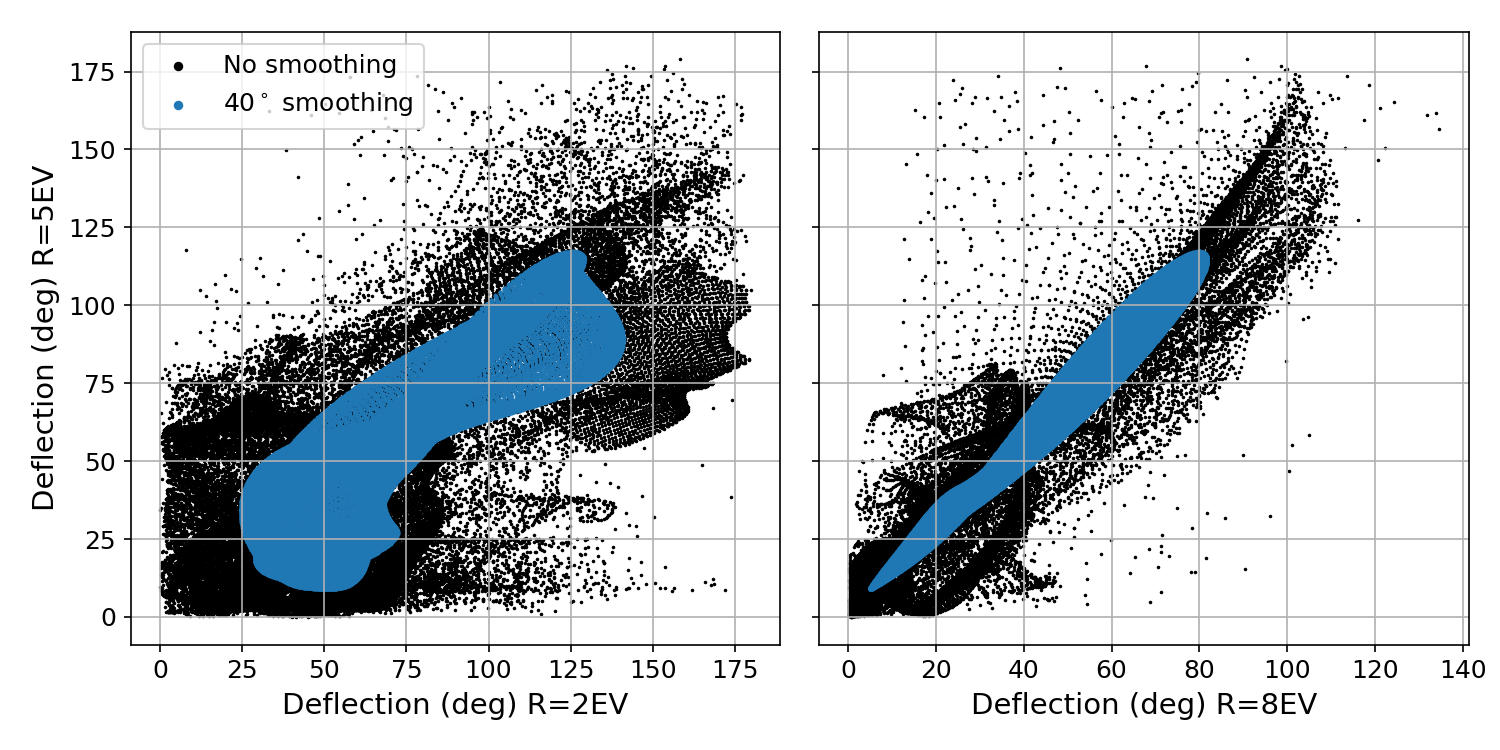}
    \caption{Comparison of GMF deflection values obtained for rigidities $R=2$ and $8$~EV with respect to the fiducial $R=5$~EV case, shown with and without angular smoothing. After smoothing, an approximately linear relation is observed between the maps, while the $R=2$~EV case displays a larger dispersion.}
    \label{fig:deflection_comparison}
\end{figure}

Fig.~\ref{fig:deflection_comparison} shows the comparison between the deflection maps obtained for $R=2$ and $R=8$~EV and those derived for the adopted fiducial value $R=5$~EV, both before and after angular smoothing. After smoothing, the correspondence between the maps is approximately linear, indicating that the large-scale structure of the GMF-induced deflections is preserved across this range of rigidities. Consequently, the expected impact on the cross-correlation analysis should follow similar trends.

\begin{figure}
    \centering
    \includegraphics[width=1\linewidth]{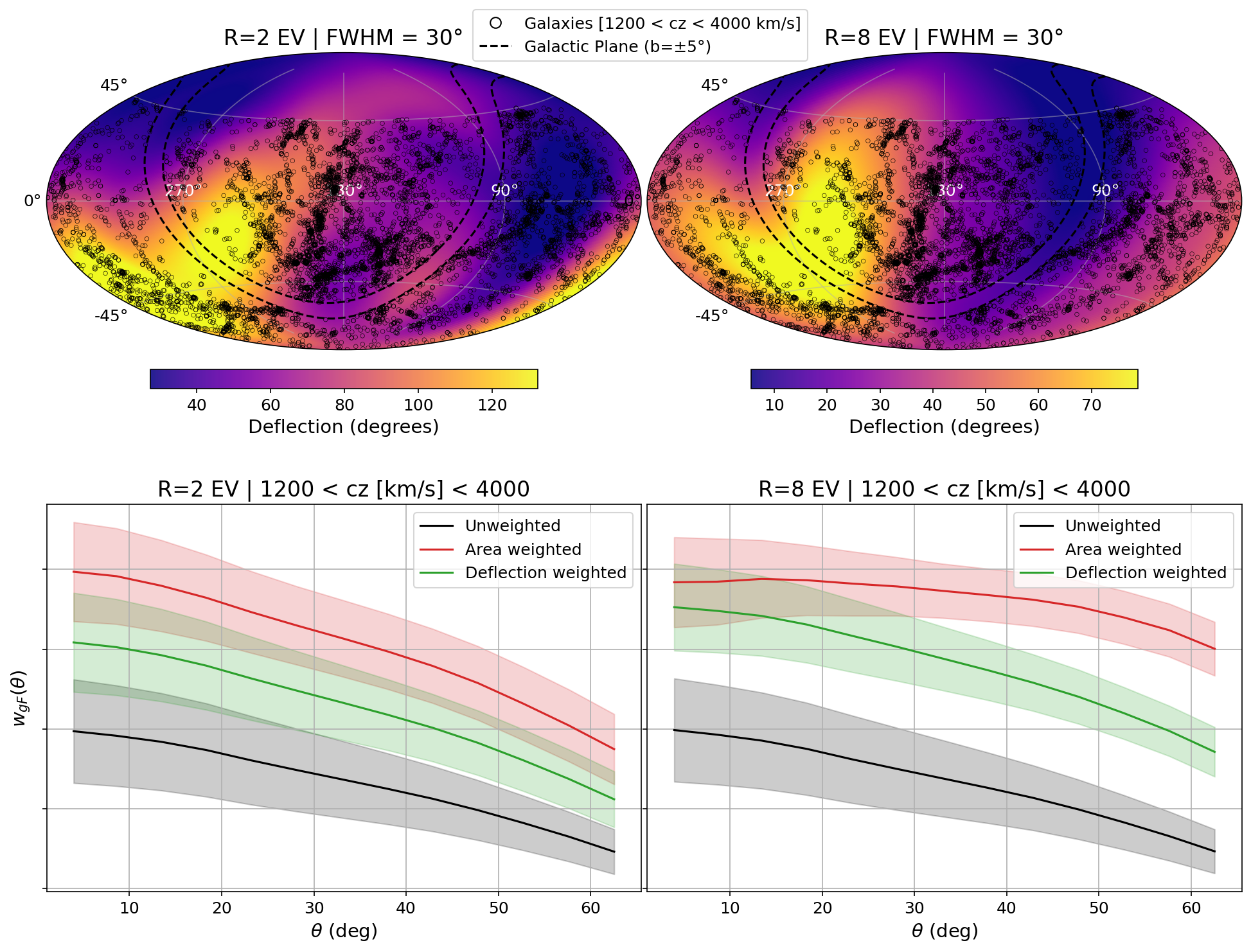}
    \caption{Top panels: Side-by-side comparison of the UF23 GMF deflection maps for rigidities $R=2$ and $8$~EV.
    Bottom panels: Cross-correlation functions obtained taking into account these rigidities. An additional weighting scheme based on the inverse deflection angle is also shown for comparison. The results follow the same qualitative trend observed for the fiducial $R=5$~EV case (Fig.~\ref{fig:correlation_weights}).}
    \label{fig:correlation_weights_R2R8}
\end{figure}

The $R=2$~EV case exhibits a larger dispersion in the deflection mapping. This behavior is expected, as such a low rigidity corresponds to either lower-energy particles or heavier nuclear species than those dominating the UHECR sample above \mbox{$E\geq8$~EeV}. Therefore, $R=2$~EV can be regarded as an extreme lower-bound scenario for the rigidity of the particles considered in this study. On the other hand, the GMF deflections with $R=8$~EV are qualitatively similar to the $R=5$~EV scenario, which can be seen by comparing Fig.~\ref{fig:skymap_defl} with the top right panel of Fig.~\ref{fig:correlation_weights_R2R8}.  

We repeated the cross-correlation analysis using these additional rigidity values. Fig.~\ref{fig:correlation_weights_R2R8} shows that the overall trend observed for $R=5$~EV (Fig.~\ref{fig:correlation_weights}) is preserved. Although the enhancement of the correlation signal is weaker for the $R=2$~EV case, the trend remains unchanged, confirming the robustness of our conclusions with respect to the assumed rigidity.

In addition, we tested an alternative weighting scheme for the galaxy catalogue based on the inverse of the deflection angle, $w_i=\theta_{i,\,\rm defl}^{-1}/\langle\theta_{\rm{defl}}^{-1} \rangle$, in contrast to the deflection-area weighting adopted in the main analysis, $w_i=\theta_{i,\,\rm defl}^{-2}/\langle\theta_{\rm{defl}}^{-2} \rangle$. While both schemes lead to an enhancement of the correlation signal, the maximum increase is obtained when weighting by the deflection area. This result supports the interpretation that the solid angle over which UHECR are spread by Galactic magnetic deflections provides the most physically appropriate weighting prescription.

\section{Additional GMF models analysis}
\label{appendix:B}

To assess the robustness of our results against uncertainties in the modeling Galactic Magnetic field (GMF), we repeat the analysis using the models of \citet[][JF12]{Jansson2012a} and \citet[][KST24]{KST2025} in addition to our fiducial \citet[][UF23]{Unger2024} model. The JF12 model was the first global GMF reconstruction simultaneously constrained by Faraday rotation measures and polarized synchrotron emission. It introduced a three-component coherent field consisting of a spiral disk, a toroidal halo, and an out-of-plane ``X'' field, together with a striated magnetic component to reproduce the observed polarized synchrotron intensity. The more recent UF23 model builds upon the JF12 framework by incorporating updated observational data, improved treatments of the thermal and cosmic-ray electron distributions, and a systematic exploration of multiple coherent field parameterizations, resulting in a family of benchmark models that better characterize the uncertainties in the large-scale GMF. In contrast, the KST24 model explicitly includes local Galactic structures neglected or masked in previous models, e.g., the Local Bubble and the Fan Region. 
By accounting for the polarized synchrotron emission arising from these nearby structures, KST24 reproduces the observational data without requiring a striated magnetic component and favors a different large-scale spiral geometry, including a larger local pitch angle. Since these models differ primarily in their treatment of the coherent field morphology and local magnetic structures, comparing the resulting UHECR deflections provides an estimate of the systematic uncertainty associated with the adopted GMF model.

\begin{figure}
    \centering
    \includegraphics[width=1\linewidth]{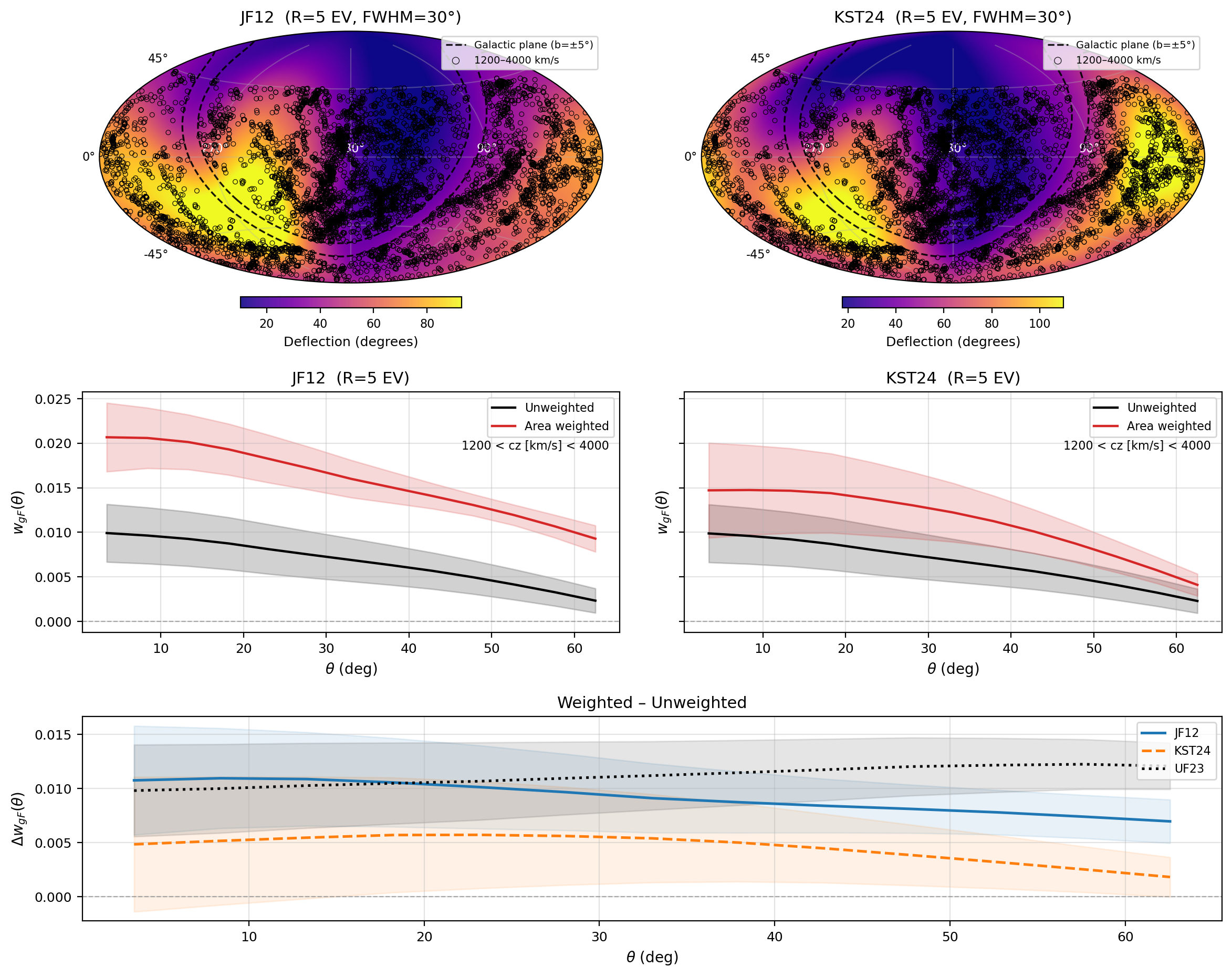}
    \caption{Cross-correlation functions obtained using two additional GMF deflection maps with \mbox{$R=5$~EV}: the \citet[][JF12]{Jansson2012a} (left panels) and the \citet[][KST24]{KST2025} models (right panels). Bottom panel shows the difference $\Delta w_{gF}$ between the weighted and unweighted correlations for each model, with the fiducial \citet[][UF23]{Unger2024} model represented with a  black dotted line.}
    \label{fig:model_comparison}
\end{figure}

Fig.~\ref{fig:model_comparison} shows the results of weighting the correlation function for the closest galaxy sample by the deflections predicted by the JF12 and KST24 models (left and right panels, respectively). The bottom panel shows the difference between the weighted and unweighted correlations for each model, with the fiducial UF23 result shown as the dotted black line. For angular scales of $\sim10^{\circ}$ or less, the weighted correlation signal increases by approximately 100\% and 50\% for the JF12 and KST24 models, respectively. The jackknife uncertainties, computed using patches that each contain a statistically significant number of galaxies, together with the intrinsic $45^{\circ}$ smoothing of the Auger map, result in conservative error bands. Nevertheless, all three GMF models consistently show that accounting for Galactic magnetic deflections through suitable weighting schemes enhances significantly the cross-correlation signal between galaxies and UHECR.

\color{black}

\section*{Acknowledgements}
We acknowledge support from CONICET, SeCyT-UNC, Foncyt and 
Secretar\'ia de Ciencia y T\'ecnica de la Universidad Nacional de San Juan. 

\section*{Data availability}
The datasets used in this article were derived from sources in the public domain.
The datasets analyzed during the current study are available from the corresponding 
author on reasonable request.

\section*{Code availability}
Code for the statistical analysis performed in this work is available from the corresponding author upon reasonable request.

\section*{Ethic declarations}
\subsection*{Competing interest}
The authors declare no competing interests.

\bibliographystyle{aasjournal}
\bibliography{biblio_xcorrCR}


\end{document}